\documentclass[12pt]{article}
\usepackage{latexsym}
\usepackage{amsfonts}
\usepackage{amsmath}
\usepackage{amsthm}
\usepackage{hyperref}

\def\d{\partial}

\def\3{\underline }
\def\5{\bar }
\def\6{\partial }
\def\7{\hat }
\def\4{\tilde }
\def\8{\bf }
\def\9{\dot }

\newcommand{\dd}[2]{\frac{\d #1}{\d #2}}

\def\half{{\frac{1}{2}}}

\def\be{\begin{eqnarray}}
\def\ee{\end{eqnarray}}
\def\beann{\begin{eqnarray*}}
\def\eeann{\end{eqnarray*}}
\def\beq{\begin{equation}}
\def\eeq{\end{equation}}
\def\ba{\begin{array}}
\def\ea{\end{array}}
\def\ben{\begin{enumerate}}
\def\een{\end{enumerate}}
\def\bea{\begin{eqnarray}}
\def\eea{\end{eqnarray}}
\def\beann{\begin{eqnarray*}}
\def\eeann{\end{eqnarray*}}
\def\beq{\begin{equation}}
\def\eeq{\end{equation}}
\def\ba{\begin{array}}
\def\ea{\end{array}}
\def\ben{\begin{enumerate}}
\def\een{\end{enumerate}}

\def\5{\bar }
\def\6{\partial }
\def\7{\hat }
\def\4{\tilde }

\def\cC{{\cal C}}

\def\cL{{\cal L}}

\def\s0#1#2{\mbox{\small{$\frac{#1}{#2}$}}}

\def\qed{\hbox{${\vcenter{\vbox{
\hrule height 0.4pt\hbox{\vrule width 0.4pt height 6pt
\kern5pt\vrule width 0.4pt}\hrule height 0.4pt}}}$}}

\begin{document}

\begin{titlepage}

\begin{flushright}
ULB-TH/04-08\\
gr-qc/0404006
\end{flushright}

\begin{centering}

\vspace{0.5cm}

{\bf{\Large Classification of surface charges for a spin $2$ field
on a curved background solution}}

\vspace{1.5cm}

{\large Glenn Barnich$^*$}

\vspace{.5cm}

Physique Th\'eorique et Math\'ematique and International Solvay Institutes,\\
Universit\'e Libre de Bruxelles,\\ Campus Plaine C.P. 231, B--1050
Bruxelles, Belgique

\vspace{1.5cm}

{\large Serge Leclercq and Philippe Spindel}

\vspace{.5cm}

M\'{e}canique \& Gravitation, Facult\'{e} des Sciences,\\ Universit\'{e} de
Mons-Hainaut \\ 20, Place du Parc, B--7000 Mons, Belgique

\vspace{1.5cm}

\end{centering}
\vspace{.5cm}

\begin{abstract}
We give an explicit proof of the result that non trivial conserved
$n-2$ forms for a spin $2$ field on a background corresponding to
a solution to Einstein's equation (with or without cosmological
constant) are characterized uniquely by the Killing vector fields
of the background.
\end{abstract}

\vspace{3.5cm}

\noindent \footnotesize{$^*$ Research Associate of the National
Fund for Scientific Research (Belgium).}

\end{titlepage}




In dimensions $n>2$, a consistent theory for a spin $2$ field
$h_{\mu\nu}$ on an Einstein space-time background can be
constructed by taking the quadratic part of the expansion of the
Einstein-Hilbert action, \bea S[g]=\frac{1}{16\pi}\int
d^nx\sqrt{|g|}(R-2\Lambda),\label{0} \eea around a given
background solution $\bar g_{\mu\nu}$. If $g_{\mu\nu}=\bar
g_{\mu\nu}+h_{\mu\nu}$, \bea S^{(2)}[h;\bar g]=\frac{1}{32\pi}\int
d^nx\,\sqrt{|\bar g|}\Big(-\half\bar D_\lambda h_{\mu\nu}\bar
D^\lambda h^{\mu\nu} + \bar D_\lambda h_{\mu\nu} \bar
D^{\mu}h^{\nu\lambda}-\bar D_\nu h \bar D_\mu h^{\mu\nu}
\nonumber\\+  \half\bar D^\lambda h\bar D_\lambda h
+\frac{\Lambda}{n-2}\,(2h_{\mu\nu}h^{\mu\nu}- h^2)\Big)\label{1}
,\eea where $\bar g_{\mu\nu}$, together with its inverse $\bar
g^{\mu\nu}$, is used to lower and raise the indices, $h=h^\mu_\mu$
and $\bar D_\mu$ denotes the associated covariant derivative. In
particular, the diffeomorphism invariance of the Einstein-Hilbert
implies the invariance of the action \eqref{1} under the gauge
transformations \bea\delta_\xi h_{\mu\nu}=L_\xi \bar
g_{\mu\nu}\qquad ,\label{2}\eea with arbitrary parameters
$\xi^\mu(x)$.

An $n-2$ form $k$ that depends on the space-time coordinates
$x^\mu$, the fields $h^a_{\mu\nu}$ and a finite number of their
derivatives\footnote{All forms that we consider below are supposed
to have this kind of dependence. In the context of the variational
bicomplex (see e.g.
\cite{Olver:1993,Saunders:1989,Anderson1991,Andersonbook} for more
details), such forms are called horizontal forms. } is conserved
if \bea dk\approx0\qquad .\eea Here $d$ is the so-called exterior
or horizontal differential defined by \[
d=dx^\lambda(\frac{\partial}{\partial
x^\lambda}+h^a_{\mu\nu,\lambda}\dd{}{h^a_{\mu\nu}}+\dots)\qquad
,\] while $\approx$ means an equality that holds "on-shell", i.e.,
for all solutions of the Euler-Lagrange equations associated to
\eqref{1}. A conserved $n-2$ form $k$ is defined to be trivial if
\bea k\approx d m\qquad ,\eea for some $n-3$ form $m$, and two
conserved $n-2$ forms are considered equivalent if they differ by
a trivial conserved $n-2$ form. By definition, these equivalence
classes are elements of $H^{n-2}_{\rm char}(d)$, the
characteristic cohomology of $d$ in form degree $n-2$ (see e.g.
\cite{vinogradov2,tsujishita,vinogradov3,Anderson1991,bryant}),
which we consider here as a vector space over $\mathbb R$.

Our aim in this note is to compute the cohomology group
$H^{n-2}_{\rm char}(d)$ for the equations of motion associated to
the action \eqref{1}. The result of this computation has been
originally presented in \cite{Anderson:1996sc}. However, since the
detailed proof is not given and does not seem to be readily
available in the literature, we will fill this gap below.

There are several reasons why the group $H^{n-2}_{\rm char}(d)$ is
relevant.

\begin{itemize}

\item In the context of the spin $2$ theory itself, conserved $n-2$ forms
give rise, upon integration over closed $n-2$ dimensional surfaces
$\cC^{n-2}$ and evaluation for a given solution, to charges that
depend only on the homology class of $\cC^{n-2}$ and the
equivalence class of $k$.

\item For space-times $g_{\mu\nu}$ that asymptotically near some
boundary approach the background metric $\bar g_{\mu\nu}$, the
appropriate generalization to the asymptotic context of the forms
$k$ can be used to define boundary charges that give information
on the properties of the space-times in the bulk (see e.g.
\cite{Abbott:1982ff,Anderson:1996sc,Barnich:2001jy}).

Alternatively, if $\bar g_{\mu\nu}$ denotes a given solution, like
a Kerr black hole for instance, and $\bar g_{\mu\nu}+h_{\mu\nu}$
an infinitesimal perturbation of this solution, then the charges
associated to $k$ are the same when evaluated on 2 closed
$n-2$-dimensional hypersurfaces that are the boundary of some
$n-1$-dimensional hypersurface. This property has been used in
\cite{Iyer:1994ys} by using the $2$-sphere at infinity and the
bifurcation $2$-sphere of a Killing horizon to derive the first
law of black hole mechanics.

Recently an improved version of the forms $k$ has been shown to be
useful in full general relativity to compute the boundary charges
on surfaces in the bulk and to interpolate between solutions that
are not infinitesimally close \cite{Barnich:2003xg}.

\item For the problem of consistently deforming the theory described
by \eqref{1} and \eqref{2} (or several copies of it), the
characteristic cohomology in form degree $n-2$ is related to
deformations that modify the algebra of the gauge transformations
\eqref{2} in a non trivial way. Indeed, a cohomological
formulation \cite{Barnich:1993vg} (see also
\cite{Stasheff:1997fe,Henneaux:1997bm}) of the problem of
consistent deformations of gauge theories \cite{Berends:1985rq}
shows that first order deformations are controlled by the local
BRST cohomology group in ghost number $0$. In turn, an important
ingredient in the computation of this group is the cohomology
group $H^n_2(\delta|d)$ of the Koszul Tate differential $\delta$,
associated to the resolution of forms pulled-back to the surface
defined by the equations of motion, modulo the horizontal
differential $d$. This last group is directly related to non
trivial deformations of the gauge algebra and can be shown
\cite{Barnich:1995db} to be isomorphic to the characteristic
cohomology in degree $n-2$.

The deformation problem for a collection of spin 2 fields on a
flat background has been considered in details in
\cite{Boulanger:2000rq}. The results obtained below are a
straightforward generalization to the curved case of the results
discussed in the proof of theorem 4.2 of \cite{Boulanger:2000rq}.

\end{itemize}

We now prove that the fields $h_{\mu\nu}$ and their derivatives
can be split into independent variables $y_A$ not constrained by
the equations of motion and  a set of variables $z^a$, in
one-to-one correspondence, at each order of the number of
derivatives involved, with the independent
  equations of motion  [see
\cite{Barnich:1995db}, section {\bf 3}]. In finding such a split,
one has to bear in mind that due to the Noether identities
associated to the gauge transformations \eqref{2}
  \[\bar D_\mu
\frac{\delta S^{(2)}}{\delta h_{\mu\nu}}\equiv 0,\] not all
equations of motions are independent. Indeed the equations defined
by \bea \{\cL_\Delta\}=\{\6_0\frac{\delta S^{(2)}}{\delta
h_{0\nu}}, \6_\rho\6_0\frac{\delta S^{(2)}}{\delta h_{0\nu}},
\6_{\rho_1}\6_{\rho_2}\6_0\frac{\delta S^{(2)}}{\delta
h_{0\nu}},\dots, \}\eea can be considered as dependent equations
that hold as a consequence of the remaining equations \bea
\{\cL_a\}=\{\frac{\delta S^{(2)}}{\delta h_{\mu\nu}},\6_k
\frac{\delta S^{(2)}}{\delta h_{0\nu}},
\6_{k_1}\6_{k_2}\frac{\delta S^{(2)}}{\delta
h_{0\nu}},\dots,\nonumber\\
\6_\rho\frac{\delta S^{(2)}}{\delta
h_{mn}},\6_{\rho_1}\6_{\rho_2}\frac{\delta S^{(2)}}{\delta
h_{mn}},\dots, \}\nonumber ,\eea where Greek indices run from $0$
to $n-1$, while Latin indices run from $1$ to $n-1$. In fact, the
equations defined by $\{\cL_a\}$ will be shown below to be all
independent by solving them for independent variables $\{z^a\}$.
At the same time, this allows us to show that the irreducible set
of gauge transformations \eqref{2} is in fact a generating set.

First we note that the equations of motion can be reexpressed as
\bea \frac{\delta S^{(2)}}{\delta
h_{\mu\nu}}=\frac{\sqrt{|g|}}{32\pi}(K^{\mu\nu}-\half\bar
g^{\mu\nu}K),\eea where \bea K^{\mu\nu}=\bar D^{\mu}\bar D^\nu
h+\bar D_\lambda\bar D^\lambda h^{\mu\nu}-2\bar D_\lambda\bar
D^{(\mu}h^{\nu)\lambda}+\frac{4\lambda}{n-2}h^{\mu\nu},\eea and
$K=\bar g_{\mu\nu}K^{\mu\nu}$. Thus the equations defined by
$\delta S^{(2)}/{\delta h_{\mu\nu}}$ and their derivatives are
equivalent to those defined by $K^{\mu\nu}$ and their derivatives.
Because the Noether identities in terms of $K^{\mu\nu}$ can still
be solved for $\6_0K^{0\nu}$, there is a similar split into
$\{K_\Delta\}$ and $\{K_a\}$ as the one into $\{\cL_\Delta\}$ and
$\{\cL_a\}$. If we assume the diagonal contravariant elements of
the metric everywhere non zero\footnote{This can always be done
locally by choosing for instance a normal geodesic coordinate
system.}, i.e., ${\bar g}^{\mu\mu}\neq 0$ for $\mu=0,\dots, n-1$,
we may introduce variables $z^a$ such that $z^a={\cal
Z}^a[K_b,y_A]$, where \bea \{z^a\}=\{&\6_0\6_0
h_{kl},\partial_\rho \6_0\6_0
h_{kl},\partial_{\rho_1}\partial_{\rho_2}\6_0\6_0
h_{kl},\dots,\nonumber\\
&\6_1\6_1 h_{0\bar\lambda},\partial_k \6_1\6_1
h_{0\bar \lambda},\partial_{k_1}\partial_{k_2}
\6_1\6_1 h_{0\bar \lambda},\dots,\nonumber\\
&\6_0\6_1 h_{22},\partial_k \6_0\6_1
h_{22},\partial_{k_1}\partial_{k_2} \6_0\6_1
h_{22}\dots\}\qquad ,\nonumber\eea with the independent variables, not
constrained by the equations of motion,  chosen as \bea
\{y_A\}=\{&h_{\mu\nu},\6_\rho h_{\mu\nu},\nonumber\\
&\6_m\6_\rho h_{k\tilde l},\6_{m_1}\6_{m_2}\6_\rho h_{k\tilde
l},\dots,\nonumber\\
&\partial_{\rho_1}\partial_{\rho_2}h_{01},\partial_{\rho_1}\partial_{\rho_2}
\partial_{\rho_3}h_{01},\dots,\nonumber\\
&\6_\rho\6_0h_{0\bar \lambda},\6_{\rho_1}\6_{\rho_2}\6_0h_{0\bar
\lambda},\dots,\nonumber\\
&\6_{\bar k}\6_lh_{0\bar \lambda},\6_{\bar k_1}\6_{\bar k_2}\6_lh_{0\bar
\lambda},\dots,\nonumber\\
&\6_{\bar k}\6_0 h_{22},\6_{\bar k_1}\6_{\bar k_2}\6_0 h_{22},\dots,\nonumber\\
&\6_{ k_1}\6_{ k_2}h_{22},\6_{ k_1}\6_{ k_2} \6_{ k_3}
h_{22},\dots\}.\nonumber
  \eea
Here barred Greek indices run from $0$ to $n-1$ without taking the
value $1$ and barred Latin indices run from $2$ to $n-1$ while
tilded Latin indices run from $1$ to $n-1$ without taking the
value $2$. One can see that the set
  of $y_A$ variables is stable with respect to the $\partial_{\bar
l}$ derivatives :   for any variable $y_A$, the derivative $\partial_{\bar
l} y_A$ belongs to the set of $y$-variable. Accordingly
  the
$y_A$ variables can be ordered according to their number $m$ of
derivatives of type $\6_{\bar l}$. Thus the "Cauchy order" of this
perturbation theory is, at most, equal to 2. Let us notice that
the same result can be obtained if we assume that globally the
background metric verifies the conditions: ${\bar g}^{01}\neq 0$
and ${\bar g}^{\bar l\bar l}\neq 0$.

After these preliminary considerations, the first step in the
computation of $H^{n-2}_{\rm char}(d)$ is the characterization of
the isomorphic group $H^n_2(\delta|d)$ (see section 6.2 of
\cite{Barnich:2000zw}). For irreducible linear gauge theory, the
latter group can be shown to be given by equivalence classes of
field dependent gauge parameters such that the corresponding gauge
transformations are antisymmetric combinations of the equations of
motion, with two parameters being equivalent if they coincide
on-shell (see \cite{Barnich:2000zw} theorem 6.7, which applies not
only to normal but also to linear irreducible gauge theories such
as the one considered here). Explicitly, we need to characterize
the vector space $Span_{\mathbb R}\,\big\{[f_\mu]\big\}$ defined
by \bea \bar D_\mu f_\nu+\bar D_\nu f_\mu\approx^{as} 0\quad
,\quad f_\mu\sim f_\mu+t_\mu\quad ,\quad t_\mu\approx 0\quad ,\eea
where $f_\mu$ depend on $x^\nu$, $h_{\lambda\rho}$ and a finite
number of their derivatives while $\approx^{as}$ denotes
antisymmetric combinations of the equations of motion in the sense
of (6.25) of \cite{Barnich:2000zw}.

In particular, $f_\mu$ satisfies $F_{\mu\nu}\equiv\bar D_\mu
f_\nu+\bar D_\nu f_\mu\approx 0$. Applying $\bar D_\lambda$ one
gets $\bar D_\lambda F_{\mu\nu}\approx 0$. Adding $\bar D_\mu
F_{\lambda\nu}\approx 0$, subtracting $\bar D_\nu
F_{\lambda\mu}\approx 0$, using $[\bar D_\alpha,\bar
D_\beta]f_\gamma=-\bar {R^\delta}_{\alpha\beta\gamma} f_\delta$
three times, gives, by using in addition the symmetry properties
of the Riemann tensor, \bea\bar D_\lambda\bar D_\mu f_\nu+\bar
{R^\sigma}_{\lambda\nu\mu}f_\sigma\approx 0\quad .\label{3}\eea

By using the freedom of adding terms that vanish when the
equations of motion hold, one can assume that $f_\mu=f_\mu(x,y)$,
where $f_\mu(x,y)$ depends on the $\6_{\bar l}$ derivatives of
$y_A$ up to some maximum order $M$. By taking $\lambda,\mu\geq 2$
in \eqref{3} and considering the dependence of $f_\nu$ on $y_A$ of
maximum order, \eqref{3} implies that $f_\nu$ cannot depend on the
variables $y_A$ of order $M$. Hence, by induction, $f_\nu$ cannot
depend on the variables $y_A$ at all, and we have shown that
$f_\mu=f_\mu(x)+t_\mu$. Furthermore, $f_\mu(x)\approx 0$ implies
$f_\mu(x)=0$, so that a solution $f_\mu(x)$ is trivial if and only
if it vanishes. For $f_\mu(x)$, the equation $\bar D_\mu
f_\nu+\bar D_\nu f_\mu\approx^{as} 0$ reduces to \bea \bar D_\mu
f_\nu+\bar D_\nu f_\mu= 0, \eea which is the Killing equation for
$\bar g_{\mu\nu}$.

{\em Hence, we have explicitly shown that $H^{n-2}_{\rm char}(d)$
is isomorphic to the vector space of Killing vectors $\tilde
\xi^\mu$ of $\bar g_{\mu\nu}$.}

The construction of associated $[k_{\tilde\xi}]$ of $H^{n-2}_{\rm
char}(d)$ is explained from a general point of view in
\cite{Barnich:2001jy}, section 3, while section 6.3.1 contains the
explicit expressions for the case at hand and recovers those
originally obtained in
\cite{Abbott:1982ff,Iyer:1994ys,Anderson:1996sc}.

\section*{Acknowledgements}
\addcontentsline{toc}{section}{Acknowledgments}

The work of G. B. is supported in part by the ``Actions de Recherche
Concert{\'e}es'' of the ``Direction de la Recherche
Scientifique-Communaut{\'e} Fran\c{c}aise de Belgique", by a
``P{\^o}le d'Attraction Interuniversitaire'' (Belgium), by
IISN-Belgium, convention 4.4505.86, by the European Commission RTN
program HPRN-CT00131, in which the authors are associated to
K.~U.~Leuven and by Proyectos FONDECYT 1970151, 7960001 (Chile).\\
S. L. and Ph. S.  acknowledge support from the Fonds National de
la Recherche Scientifique through an F.R.F.C. grant.


\providecommand{\href}[2]{#2}\begingroup\raggedright\endgroup

\end{document}